\documentclass{article}
\usepackage{spconf,amsmath,graphicx}
\usepackage{cases}
\usepackage{multirow, tabularx}
\usepackage{hyperref}
\usepackage{xfrac} 
\usepackage{float}
\usepackage{algorithm}
\usepackage{algpseudocode}
\usepackage{enumitem}
\usepackage{pdfpages}
\usepackage{amssymb}
\usepackage{pifont}

\newcolumntype{C}{>{\centering\arraybackslash}X}
\newcommand{\xmark}{\ding{55}}%
\newcommand{\cmark}{\ding{51}}%
\newcommand\blfootnote[1]{%
  \begingroup
  \renewcommand\thefootnote{}\footnote{#1}%
  \addtocounter{footnote}{-1}%
  \endgroup
}
\restylefloat{table}
\hypersetup{
    urlcolor=blue
}

\title{M3FPolypSegNet: Segmentation Network with Multi-frequency Feature Fusion for Polyp Localization in Colonoscopy Images}
\name{Ju-Hyeon Nam, Seo-Hyeong Park, Nur Suriza Syazwany, Yerim Jung, Yu-Han Im and Sang-Chul Lee}
\address{Department of Computer Science and Engineering, Inha University, Incheon, Republic of Korea}
\begin{document}
%
\maketitle
\begin{abstract}
Polyp segmentation is crucial for preventing colorectal cancer a common type of cancer. Deep learning has been used to segment polyps automatically, which reduces the risk of misdiagnosis. Localizing small polyps in colonoscopy images is challenging because of its complex characteristics, such as color, occlusion, and various shapes of polyps. To address this challenge, a novel frequency-based fully convolutional neural network, \textit{Multi-Frequency Feature Fusion Polyp Segmentation Network (M3FPolypSegNet)} was proposed to decompose the input image into low/high/full-frequency components to use the characteristics of each component. We used three independent multi-frequency encoders to map multiple input images into a high-dimensional feature space. In the \textit{Frequency-ASPP Scalable Attention Module (F-ASPP SAM)}, ASPP was applied between each frequency component to preserve scale information. Subsequently, scalable attention was applied to emphasize polyp regions in a high-dimensional feature space. Finally, we designed three multi-task learning (i.e., region, edge, and distance) in four decoder blocks to learn the structural characteristics of the region. The proposed model outperformed various segmentation models with performance gains of 6.92\% and 7.52\% on average for all metrics on CVC-ClinicDB and BKAI-IGH-NeoPolyp, respectively.
\end{abstract}

\blfootnote{This work was supported in part by the National Research Foundation of Korea (NRF) under Grant NRF-2021R1A2C2010893 and in part by Institute of Information \& communications Technology Planning \& Evaluation (IITP) grant funded by the Korea government(MSIT) (No.RS-2022-00155915, Artificial Intelligence Convergence Innovation Human Resources Development (Inha University).}
\section{Introduction} \vspace{-0.3cm}
\label{sec:intro}

Colorectal Cancer is one of the most common types of cancer worldwide \cite{bernal2017comparative}. Because colorectal cancer typically starts as polyps and progresses to cancer, medical experts recommend regular colonoscopies for the early detection of polyps. However, manual polyp detection is not highly accurate, because of dependence on the ability of doctors and the limitations of colonoscopy equipment, which results in a decreasing survival rate.

With the advancement of deep learning \cite{long2015fully,chen2017rethinking}, automatic polyp segmentation has been developed rapidly to reduce misdiagnosis resulting from overworked doctors and obsolete equipment. U-Net \cite{ronneberger2015u} has been widely adopted in polyp segmentation tasks because of its remarkable performance in biomedical image segmentation. In U-Net++ \cite{zhou2019unet++}, the ensemble nested U-Net of various depths and deep supervision are used. ResNet++ \cite{jha2019resunet++} is focused on attention mechanisms and multi-scale feature extraction. In PraNet \cite{fan2020pranet}, reverse attention is used to clarify the relationship between areas and boundary cues to mitigate misaligned prediction.

The localization of small polyps in polyp segmentation is challenging because of complex structures such as, colors, occlusion, and various shapes of polyps and affects model performance. Frequency-based methods exhibit considerable potential for image segmentation. In FRCU-Net \cite{azad2021deep}, the Laplacian pyramid and Frequency Re-Calibration module that implement frequency attention to the basic U-Net architecture was applied. By contrast, in FDA \cite{yang2020fda}, discrete Fourier transform (DFT) is applied to each image, replacing the low-frequency component of the target image with the source image and the source image with the target style is reconstructed through inverse DFT.

\begin{figure*}[t]
    \centering
    \includegraphics[width=0.95\textwidth]{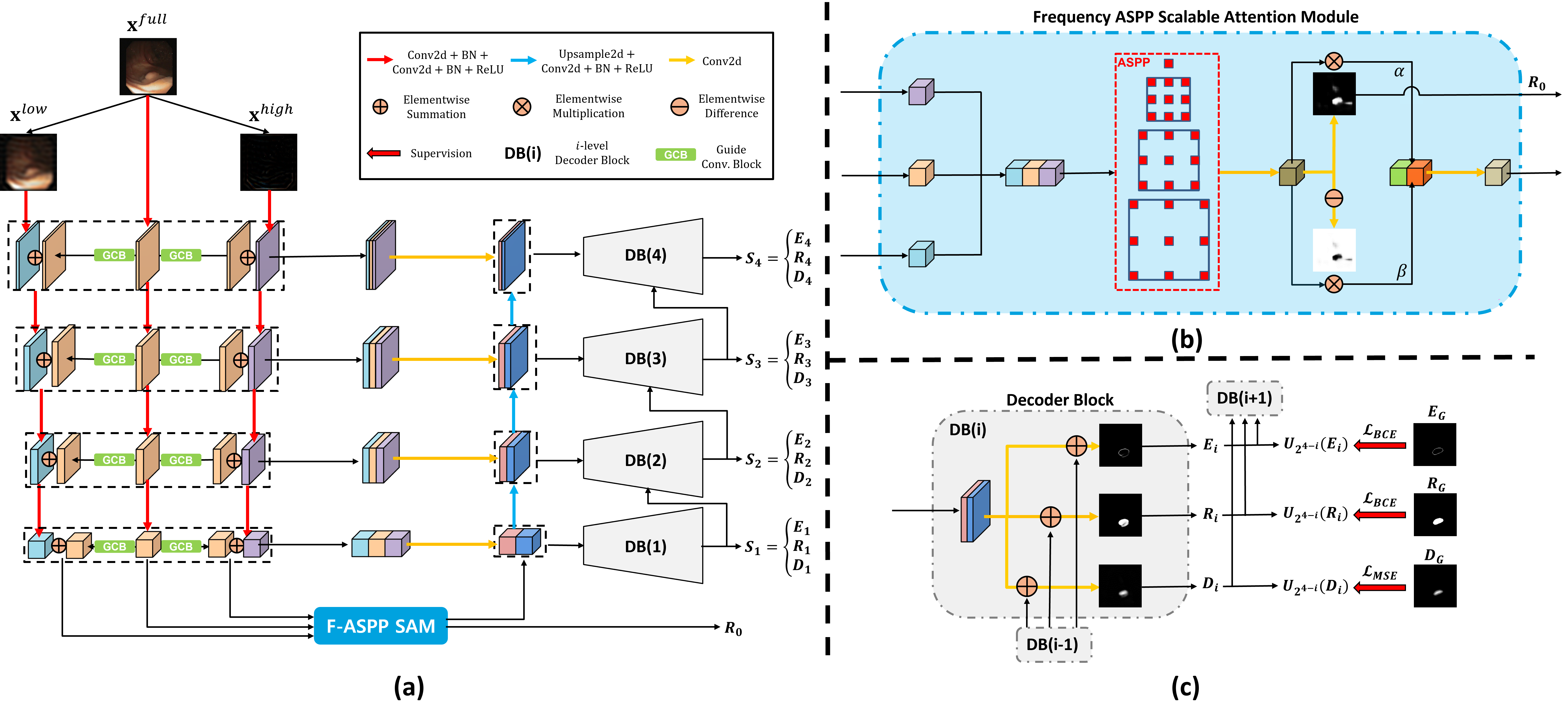} \vspace{-0.55cm}
    \caption{The proposed M3FPolypSegNet architecture. (a) Overall block diagram of our network, (b) Overview of F-ASPP SAM, and (c) Training procedure of $i$-level decoder block $\textbf{DB(i)}$ with multi-task learning and $S_{i} = \{R_{i}, E_{i}, D_{i}\}$ is multiple output from each decoder block $\textbf{DB(i)}$ for $i = 1, 2, 3, 4$.} \vspace{-0.6cm}
    \label{fig:M3FPolypSegNet}
\end{figure*}

In this paper, we propose a novel frequency-based fully convolutional neural network (FCNN), Multi-Frequency Feature Fusion Polyp Segmentation Network (M3FPolypSegNet). M3F PolypSegNet extracts feature maps by decomposing the input image into low/full/high-frequency to learn unique characteristics. The Frequency-ASPP Scalable Attention Module (F-ASPP SAM) combines the modified ASPP and attention modules to preserve scale information and emphasize polyp regions in a high-dimensional feature space. Finally, a multi-task (i.e., region, edge, and distance) loss is designed for learning the structural characteristics of the polyp region during training. The proposed model outperformed various segmentation models with performance gains of 6.92\% and 7.52\% on average for all metrics on CVC-ClinicDB and BKAI-IGH-NeoPolyp, respectively. The contributions of the study are as follows:

\begin{itemize}
\item We propose a novel polyp segmentation model (M3F PolypSegNet) based on a multi-frequency encoder and a single-decoder architecture that utilizes unique characteristics for each frequency component. \vspace{-0.25cm}
\item F-ASPP SAM introduces trainable parameters between the foreground/background attention of frequency and scale to prevent information loss during the gradual upsampling of the decoder block. Furthermore, the vanishing gradient problem was prevented by performing multi-task deep supervision training in each decoder block. \vspace{-0.25cm}
\item We experimentally achieved state-of-the-art performance in various evaluation metrics when comparing various polyp image segmentation models on two datasets (CVC-ClinicDB and BKAI-IGH-NeoPolyp). \vspace{-0.25cm}
\end{itemize}

\section{Method} \vspace{-0.3cm}
\label{sec:method}
M3FPolypSegNet is a novel polyp segmentation architecture with multiple encoders and a single decoder-based FCNN for end-to-end training. Our model consists of three primary components, namely multi-frequency encoder, frequency-ASPP scalable attention module, and single decoder with multi-task learning. Figure \ref{fig:M3FPolypSegNet} (a) display the overall architecture of M3FPolypSegNet.

\subsection{Multi-Frequency Encoder} \vspace{-0.3cm}
We were motivated by \cite{wang2020high}, in which the low/high-frequency components exhibit distinct cues that contain the approximate location and details from the low/high-frequency polyp image, respectively. Therefore, our motivation is that if latent feature maps of images are extracted and combined with various frequencies, a multi-modality perspective can be used. For this purpose, we applied DFT to transform the input image in the frequency domain.

Let $\mathbf{x}^{full} \in \mathbb{R}^{H \times W \times 3}$ be the input image where $H$ and $W$ are the height and width of the input image, respectively. First, we transform the input image into the frequency domain using DFT as follows:

\begin{equation}
\hat{\mathbf{x}}^{full} = \mathcal{F}\{\mathbf{x}^{full}\} \in \mathbb{C}^{H \times W \times 3}    
\end{equation}

where $\mathcal{F}\{ \cdot \}$ denotes the DFT with a shift operator that multiplies $(-1)^{u + v}$ in the frequency domain to translate the DC component to the center of the image where $u$ and $v$ are the frequency index. The low/high-frequency components based on the proportion of the total power spectrum of the image are defined. Let $T$ and $P$ be the total and partial power spectrum of the image, respectively. For a power spectrum ratio $0 \le r \le 1$, we define the low-frequency component satisfying $P / T \le r$. After finding the maximum frequency index $(u_{max}, v_{max})$ satisfying the condition, we define the low-frequency pass mask $M_{low}$  as follows:

\begin{equation}
M_{low}(u, v) = \begin{cases} 1 \text{ if } u^{2} + v^{2} \le  R^{2} \\ 0 \text{ if Otherwise} \end{cases}   
\end{equation}

where $R^{2} = u_{max}^{2} + v_{max}^{2}$. We decompose $\hat{\mathbf{x}}$ into the low/high-frequency components by two binary masks, namely $M_{low}$ and $1 - M_{low}$ to create a new input image as follows:

\begin{equation}
    \begin{cases}
    &\mathbf{x}^{low} = \mathcal{F}^{-1}\{\mathbf{x}^{full} \otimes M_{low}\} \\ 
    &\mathbf{x}^{high} = \mathcal{F}^{-1}\{\mathbf{x}^{full} \otimes (1 - M_{low})\} 
    \end{cases}
\end{equation}

where $\mathcal{F}^{-1}\{ \cdot \}$ and $\otimes$ represent the inverse DFT with a shift operator that multiplies $(-1)^{x + y}$ and element-wise multiplication, respectively. We design an encoder-decoder architecture similar to U-Net \cite{ronneberger2015u} in this paper. The encoder consists of four blocks with a convolutional layer, batch normalization, and ReLU activation function. First, a high-dimensional feature map was extracted from a full-frequency image for each $i = 1, 2, 3, 4$ as follows:

\begin{equation}
\mathbf{x}_{i}^{full} = e_{i}^{full}\left(\mathbf{x}_{i - 1}^{full}\right)
\end{equation}

where $\mathbf{x}_{0}^{full} = \mathbf{x}^{full}$ , and $e_{i} (\cdot)$ is $i$-th encoder block. The multi-frequency encoder has higher representation power than a single encoder by training multiple encoders from images in various frequency components. However, $\mathbf{x}^{low}$ and $\mathbf{x}^{high}$ exhibit information loss, because specific frequency ranges are completely removed. To solve this problem, we add a residual connection, Guided Convolution Block (GCB), from a full-frequency encoder into low/high-frequency encoders while extracting high-dimensional feature maps as follows:

\begin{equation}
    \mathbf{x}_{i}^{z} = e_{i}^{z} \left( \mathbf{x}_{i - 1}^{z} \right) + GCB\left(\mathbf{x}_{i}^{full}\right)
\end{equation}

where $z = low, high$ and $GCB(\cdot)$ consists of $1 \times 1$ convolution and ReLU activation function. Because the three inputs have distinct characteristics, we use three independent encoders that do not share parameters. Through this method, each encoder extracts feature maps corresponding to each frequency component while minimizing information loss.

\subsection{Frequency-ASPP Scalable Attention Module} \vspace{-0.3cm}
In the F-ASPP SAM, heterogeneous feature maps are fused to enhance polyp regions. The architecture of F-ASPP SAM is displayed in Figure \ref{fig:M3FPolypSegNet} (b). First, the three feature maps are concatenated as $\mathbf{X} = \left[ \mathbf{x}_{4}^{low}, \mathbf{x}_{4}^{full}, \mathbf{x}_{4}^{high} \right] \in \mathbb{R}^{H/16, W/16, 3C_{4}}$. Convolutional layers with various atrous rates are then used for efficient multi-frequency and scale fusion. In this paper, after modifying the original ASPP architecture, the four branches use convolution operations with various atrous rates to extract feature maps and then sum them.

However, this method cannot capture polyp regions, and the results tend to be scattered. Therefore, we simultaneously applied foreground/background attention and concatenated them to preserve the information of polyp regions during progressive decoding. We introduced two trainable parameters ($\alpha$ and $\beta$) to determine two attention ratios as follows:

\begin{equation}
O = C_{3 \times 3}([\left(\alpha g(f(\mathbf{X}))\right)\mathbf{X}, \left(\beta (1 - g(f(\mathbf{X})))\right)\mathbf{X}])
\end{equation}

where $f(\cdot)$ and $g(\cdot)$ are four-branch ASPP module which we modified from original module \cite{chen2018encoder} and $1 \times 1$ convolution block, respectively. The result of the concatenation goes through a $3 \times 3$ convolutional layer and continues with the decoding.

\subsection{Training and Inference Process} \vspace{-0.3cm}
By applying deep supervision to each block of the decoder, we obtained four additional outputs $R_{0}, S_{1}, S_{2}, S_{3}$ and the final output $S_{4}$. We denote that $S_{i} = \{R_{i}, E_{i}, D_{i}\}$ is multiple outputs from each decoder block, $\textbf{DB(i)}$, that performs multi-task learning for each $i = 1, 2, 3, 4$. At this stage, each task prediction is upsampled to the same size as the ground truth to calculate the loss function. The edge ground truth, $E_{G}$, is obtained by applying the anisotropic Sobel edge detection filter from $R_{G}$. The distance map ground truth, $D_{G}$, is obtained from $R_{G}$ by applying a distance transform and normalizing the distances from pixels in the region to edges between 0 and 1. The loss function for the $i$th decoder block, $\textbf{DB(i)}$, is computed as follows:

\begin{equation}
\begin{split}
    \mathcal{L}^{i} = \mathcal{L}_{BCE}(U_{2^{4 - i}}(R_{i}), R_{G}) + \mathcal{L}_{BCE}(U_{2^{4 - i}}(E_{i}), E_{G}) \\ + \mathcal{L}_{MSE}((U_{2^{4 - i}}(D_{i}), D_{G})
\end{split}
\end{equation}

\begin{table*}[tb]
    \centering
    \footnotesize	
    \begin{tabularx}{\textwidth}{c|c|ccccc|ccccc}
    \hline
    \multirow{2}{*}{Method} & \multirow{2}{*}{Parameters} & \multicolumn{5}{c|}{CVC-ClinicDB \cite{bernal2015wm}} & \multicolumn{5}{c}{BKAI-IGH-NeoPolyps \cite{ngoc2021neounet}} \\
     & & Acc & F1-Score & Recall & Precision & IoU & Acc & F1-Score & Recall & Precision & IoU \\
     \hline
    \multicolumn{1}{c|}{FCN8s \cite{long2015fully}} & 18.64M & 0.9723 & 0.8015 & 0.8175 & 0.8299 & 0.7285 & 0.9659 & 0.7234 & 0.8399 & 0.7092 & 0.6726 \\
    \multicolumn{1}{c|}{DeepLabV3+ \cite{chen2018encoder}} & 59.34M & 0.9791 & 0.8373 & 0.8306 & 0.8736 & 0.7844 & 0.9806 & 0.8822 & 0.9066 & 0.8881 & 0.8422 \\
    \multicolumn{1}{c|}{SegNet \cite{badrinarayanan2017segnet}} & 16.50M & 0.9631 & 0.5787 & 0.6862 & 0.5707 & 0.5484 & 0.9501 & 0.6440 & 0.7238 & 0.6374 & 0.6115 \\
    \multicolumn{1}{c|}{U-Net \cite{ronneberger2015u}} & 34.53M & 0.9792 & 0.8585 & 0.8635 & 0.8962 & 0.7985 & 0.9842 & 0.9052 & 0.9217 & 0.9138 & 0.8696 \\
    \multicolumn{1}{c|}{U-Net++ \cite{zhou2018unet++}} & 36.63M & 0.9827 & 0.8817 & \textit{0.8897} & 0.9157 & 0.8257 & 0.9861 & 0.9178 & \textbf{0.9515} & 0.9194 & 0.8878 \\
    \multicolumn{1}{c|}{CENet \cite{gu2019net}} & 29.00M & \textit{0.9845} & \textit{0.8854} & 0.8699 & \textbf{0.9249} & \textit{0.8370} & 0.9869 & 0.9133 & 0.9285 & 0.9161 & 0.8790 \\
    \multicolumn{1}{c|}{ResU-Net \cite{zhang2018road}} & 10.81M & 0.9522 & 0.7812 & 0.7563 & 0.8720 & 0.7022 & 0.9791 & 0.8860 & 0.9222 & 0.8878 & 0.8362 \\
    \multicolumn{1}{c|}{ResU-Net++ \cite{jha2019resunet++}} & 14.48M & 0.9684 & 0.7643 & 0.7907 & 0.7948 & 0.7026 & 0.9476 & 0.7846 & 0.8887 & 0.7611 & 0.7152 \\
    \multicolumn{1}{c|}{PraNet \cite{fan2020pranet}} & 32.55M & 0.9797 & 0.8748 & 0.8654 & \textit{0.9244} & 0.8248 & \textbf{0.9918} & \textit{0.9370} & \textit{0.9461} & \textit{0.9395} & \textit{0.9081} \\
    \multicolumn{1}{c|}{\textbf{M3FPolySegNet (Ours)}} & 22.39M & \textbf{0.9883} & \textbf{0.8937} & \textbf{0.9015} & 0.9062 & \textbf{0.8507} & \textit{0.9891} & \textbf{0.9399} & 0.9441 & \textbf{0.9450} & \textbf{0.9147} \\
    \hline
    \end{tabularx}
    \caption{Experiment results on the CVC-ClinicDB and BKAI-IGH-NeoPolyps datasets. \textbf{Bold}  and \textit{italic} denote best and second-best performance, respectively.} \vspace{-0.5cm}
    \label{tab:summary}
\end{table*}

where $U_{2^{4 - i}}(\cdot)$ is bi-linear interpolation with a $2^{4 - i}$ scale factor. Finally, the total loss function in M3FPolypSegNet is $\mathcal{L}_{total} = \mathcal{L}_{BCE}(R_{0}, R_{G}) + \sum_{i = 1}^{4} \mathcal{L}^{i}$. Additionally, the final prediction map $R_{final}$ can be obtained by applying the sigmoid function to $R_{4} \in S_{4}$.

\section{Experimental Results} \vspace{-0.3cm}
\label{sec:results}

\subsection{Experimental Settings and Implementation Details} \vspace{-0.3cm}
We implemented M3FPolypSegNet in Pytorch 1.11 and Python 3.8, and used two datasets (CVC-ClincDB \cite{bernal2015wm} and BKAI-IGH-NeoPolyp \cite{ngoc2021neounet}) for training and evaluation. All input images were resized at the same resolution of $256 \times 256$. We compared the proposed M3FPolypSegNet with ten existing segmentation networks (FCN8s \cite{long2015fully}, DeepLabv3+ \cite{chen2018encoder}, SegNet \cite{badrinarayanan2017segnet}, U-Net \cite{ronneberger2015u}, U-Net++ \cite{zhou2018unet++}, CENet \cite{gu2019net}, ResU-Net \cite{zhang2018road}, ResU-Net++ \cite{jha2019resunet++}, PraNet \cite{fan2020pranet}). We optimized parameters using the Adam optimizer in an end-to-end approach. The initial learning rate started from $10^{-4}$ and decreased to $10^{-6}$ by using the cosine annealing learning rate scheduler, and the training settings were set to a batch size of 16 and epochs of 200. During the training phase, a random horizontal flipping with a probability of 50\% and a random non-extended rotation between $-5^{\circ}$ and $5^{\circ}$ were applied. We use five representative segmentation metrics (pixel accuracy, precision, recall, F1-Score, and IoU) for comparison. We fixed $r = 0.5$ to equally set the importance between low/high-frequency\footnote[1]{The code is available in our \href{https://github.com/ICIP2023/M3FPolypSegNet.git}{M3FPolypSegNet.git}}.

\begin{figure}[t]
    \centering
    \includegraphics[width=0.45\textwidth]{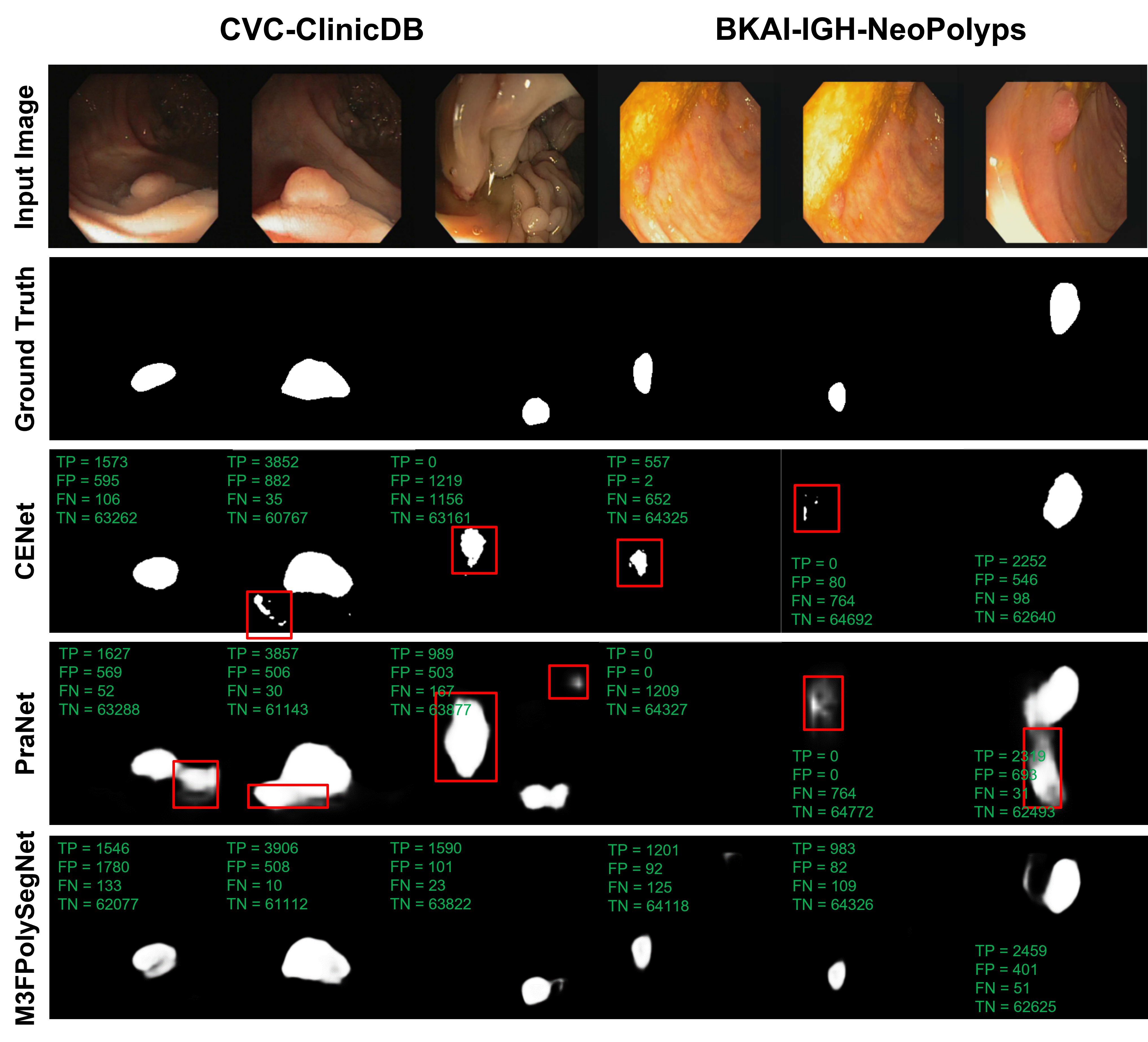} \vspace{-0.55cm}
    \caption{Qualitative results of PraNet \cite{fan2020pranet}, CENet \cite{gu2019net} and M3FPolypSegNet for each input image. Red boxes indicate misdiagnosis of the input images.} \vspace{-0.6cm}
    \label{fig:Qualitative}
\end{figure}
 
\subsection{Results Analysis} \vspace{-0.3cm}
Table \ref{tab:summary} summarizes experiment results. As presented in Table \ref{tab:summary}, M3FPolypSegNet outperformed on all metrics but exhibited a much higher recall on the CVC-ClnicDB dataset. In particular, IoU is improved by approximately 2.5\%, and 1.3\% compared with PraNet and CENet, respectively. We introduce two trainable attention ratios ($\alpha$ and $\beta$) between foreground and background in F-ASPP SAM. Therefore, M3FPolypSegNet's prediction mask can observe higher details of polyp edges compared to PraNet \cite{fan2020pranet}, which only uses existing reverse attention. Furthermore, M3FPolySegNet obtained higher performance than PraNet even if approximately 10M lower number of parameters. In Figure \ref{fig:Qualitative}, we observe the qualitative results of each method. In colonoscopy image, a single lighting source makes difficulties that edges of polyps are ambiguous. Our model leverages the high-frequency components to increase invariance and perform specialized segmentation on colonoscopy images. High-frequency components tend to reveal detailed information, such as the edges of polyps, and by utilizing this information, the accuracy and performance of polyp segmentation can be improved. \vspace{-0.35cm}

\subsection{Ablation Study} \vspace{-0.35cm}
In this section, we measure the performance of each component of M3FPolypSegNet; Frequency(FD), GCB, Multi-Task Learning (MTL) and Frequency-ASPP Scalable Atention Module (F-ASPP SAM), separately on two datasets to gain a deeper understanding of our model. The results are summarized in the Table \ref{tab:ablation}.

First, when FD and GCB are applied to U-Net, performance degradation occurs on CVC-ClinicDB and BKAI. However, after MTL is applied for each decoder, the performance is improved by 2.97\% and 1.24\% compared with U-Net. Table \ref{tab:ablation} reveals that keeping MTL and removing GCB results in performance degradation of 3.49\% and 4.04\%, respectively. This result indicates that GCB supplement information loss due to FD and gain performance improvement. Finally, when F-ASPP SA is added, the performance of both datasets is improved by a large margin. It can be seen that CVC-ClinicDB and BKAI-IGH-NeoPolyp improve from 0.8282 and 0.882 to 0.8507 and 0.9147, respectively.

\begin{table}[t]
    \centering
    \footnotesize
    \begin{tabularx}{0.48\textwidth}{c|c|c|c|c|c}
    \hline
    FD       & GCB    & MTL    & F-ASPP SAM & CVC-ClincDB & BKAI \\
    \hline
    \textcolor{red}{\xmark}   & \textcolor{red}{\xmark} & \textcolor{red}{\xmark} & \textcolor{red}{\xmark}     & 0.7985 & 0.8696    \\
    \textcolor{blue}{\cmark}   & \textcolor{blue}{\cmark} & \textcolor{red}{\xmark} & \textcolor{red}{\xmark}     & 0.7909 & 0.8684    \\
    \textcolor{blue}{\cmark}   & \textcolor{red}{\xmark} & \textcolor{blue}{\cmark} & \textcolor{red}{\xmark}     & 0.8103 & 0.8771    \\
    \textcolor{blue}{\cmark}   & \textcolor{blue}{\cmark} & \textcolor{blue}{\cmark} & \textcolor{red}{\xmark}     & \textit{0.8282} & \textit{0.8820}    \\
    \textcolor{blue}{\cmark}   & \textcolor{blue}{\cmark} & \textcolor{blue}{\cmark} & \textcolor{blue}{\cmark}     & \textbf{0.8507} & \textbf{0.9147}    \\
    \hline
    \end{tabularx}
    \caption{Ablation study on the CVC-ClinicDB and BKAI-IGH-NeoPolyps (BKAI) datasets. \textbf{Bold}  and \textit{italic} denote best and second-best performance, respectively.} \vspace{-0.5cm}
    \label{tab:ablation}
\end{table}

\section{Conclusion} \vspace{-0.35cm}

We propose M3FPolypSegNet, a polyp segmentation model based on frequency-domain automated colonoscopy images. Experiment results revealed that M3FPolypSegNet exhibits higher learning and evaluation capabilities than existing polyp segmentation models (CVC-ClinicDB: $>6\%$ \& BKAI-IGH-NeoPolyp: $>7\%$). In particular, the power spectrum-based frequency decomposition technique and multi-frequency-based feature fusion method enable high-performance improvements by preventing spatial information loss during training. Furthermore, we demonstrated that the proposed model can be applied to various datasets because it does not require any initialization techniques or post-processing techniques. We conducted additional research in areas such as various topics related to biomedical images (brain tumor segmentation, liver segmentation, etc.), rather than restricting M3FPolypSegNet to polyp segmentation task.

\bibliographystyle{IEEEbib}
\bibliography{main}

\end{document}